\documentclass[12pt]{article}
\newcommand{\beq}{\begin{equation}}
\newcommand{\eeq}{\end{equation}}
\newcommand{\bra}{\begin{array}}
\newcommand{\era}{\end{array}}

\newcommand{\te}{\theta}
\newcommand{\al}{\alpha}

\newcommand{\de}{\delta}
\newcommand{\da}{\dagger}

\newcommand{\om}{\omega}

\usepackage{latexsym}
\author{J. Douari \footnote{douari@sun.ac.za} \\
\small\it Stellenbosch Institute for Advanced Study, Private Bag X1,\rm\\
\small\it Matieland, Stellenbosch, 7601, South Africa\rm }
\title{New Construction of Exotic Particles Algebra and Noncommutative Geometry}
\begin{document}
\maketitle
\vspace*{0.5cm}
PACS: 03.65.Fd, 03.65.-w, 03.70.+k
\vskip1cm
Keywords: Planar System, Non-Commuting coordinates, Exotic particles algebra.
\vspace*{1cm}
\section*{Abstract}
\hspace{.3in}This letter establishes a procedure which can determine an algebra of exotic particles obeying fractional statistics and living in two-dimensional space using a non-commuting coordinates.
\section{Introduction}
\hspace{.3in}The goal of this work is mainly to obtain an algebra describing anyons. These particles are also known as excitations, quasi-particles or exotic particles; i.e. fermions (bosons) carrying odd (even) number of elementary magnetic flux quanta \cite{an}. They are living in two-dimensional space as composite particles having arbitrary spin, and they are characterized by fractional statistics which is interpolating between bosonic statistics and fermionic one. Several works were done to find out their quantum theory and its right model is not yet reached.\\

The construction of the exotic particles algebra we give in this letter is different from the one introduced by Lerda and Sciuto in the ref. \cite{as}. As result, the obtained symmetry is interpolating between bosonic and deformed fermionic algebras. The procedure is based on considering a non-commutative geometry depending on the statistical parameter characterizing the system. First we review the non-commutative geometry as known in the literature which is one of the approaches to describe the non-commutative gauge theories which are the topic of the most recent interest \cite{ncg}. They are described through two ways: by considering the standard commutative gauge action and re-interpreting any product of arbitrary fields in terms of the Moyal product, and by re-interpreting the fields as operators in the Hilbert space which provides for a representation of the fundamental algebra that defines the non-commutative geometry \cite{nce}. Then we suggest that the commutation relations of coordinates should be depending on statistical parameters $\nu$ characterizing exotic particles. The obtained non-commutative geometry is used to construct an annihilation and creation operators generating the exotic particles algebra. An interesting remark is that the latter symmetry has two extremes the bosonic algebra for $\nu=0$ and deformed fermionic algebra for $\nu=1$ which means that the discussed planar system is gotten from bosons as origin and does not have any thing to do with fermions.\\

The letter is organized as follows: In section 2, we briefly review the origin of anyons and its statistics. The section 3 will be devoted to construct an exotic particles algebra basing on the non-commuting coordinates which generate the algebra underlying the non-commutative geometry depending on the statistical parameter. In section 4, we conclude.

\section{Exotic Particles}
\hspace{.3in}Let us start by recalling the origin of anyons \cite{an}. These particles are known as excitations in two-dimensional space obey intermediate statistics that interpolate between bosonic and fermionic statistics. It was proved, in fractional quantum Hall effect case, that electrons don't act as a gas like they do in normal metals. They have condensed to form a new type of quantum fluid. Because electrons have a great reluctance to condense, each must first capture and combine with a quantum unit of magnetic flux. This was the proposition of Robert Laughlin \cite{qhe}. The unique property of his quantum fluid is that if one electron is added, the fluid is excited creating quasi-particles that have a fraction of the charge of an electron. These are not normal particles, but just the coordinated motion of several electrons in the fluid called anyons.\\

Thus, to see how come anyons as a theory, firstly, as known the configuration space $M^d _N$ of $N$ identical particles in $d$-dimensional space $(\Re^d )^N$  is given as follows
$$M^d _N =\frac{(\Re^d )^N -\Delta}{S_N }$$
by removing the diagonal $\Delta$ defined by the set $$\Delta=\lbrace (x_1 ,...,x_N )\in (\Re^d )^N / x_i =x_j \rbrace$$ such that $x_i =x_j$ for at least one pair.  Here we can imagine that there is a hard core interaction between particles keeping them apart. Then we identify the elements of the configuration space $(x_1 ,...,x_N )$ and $(x_{\pi(1)} ,...,x_{\pi(N)} )$, for any element $\pi$ of the symmetry group $S_N$ since particles are identical.\\

In the special case two-dimensional space with two identical particles, the configuration space $M^2 _2$ 
$$
M^2 _2 =\Re^2 \times\lbrace \mbox{cone without the tip} \rbrace 
$$
is infinitely connected. It is constructed by replacing the coordinates $x_1$ and $x_2$ by the center of mass coordinate $X=\frac{x_1 +x_2 }{2}$ and the relative coordinate $x=x_1 - x_2$ and by removing the diagonal $x_1 = x_2$ means leaving out the origin of the $x$-plane and "modding" by $S_2$ means identifying $x$ and $-x$. Then, the resulting construction is the surface of a cone with the tip $x=0$ excluded. Consequently, any closed loop on the mantle of the cone encircling the tip can not be shrunk to a point. Thus  $M^2 _2$ is multiply connected.\\

Secondly, the connection to statistics comes through recognizing that the class of closed loops $\sigma_i$ corresponds to an interchange of particles $i$ and $i+1$ (Figure 1). In $d\ge 3$, these loops can be deformed into each other; e.g. by rotating the loop around a diameter of a sphere, then $\sigma_i =\sigma_i ^{-1}$. In $d=2$, this can be done in two homotopically inequivalent ways which can be represented by the loop $C_i C_{i+1}$ where the two particles move either counterclockwise (corresponding to $\sigma_i $) or clockwise (corresponding to $\sigma_i ^{-1}$) interchanging their places, so $\sigma_i \ne\sigma_i ^{-1}$ and they are elements of the braid group $B_N$. The latter condition is the difference between $B_N$ and $S_N$.

\begin{center}

\setlength{\unitlength}{.2in}
\begin{picture}(20,6)

\put(3.13,0.07){\oval(6,5)[t]}

\put(3.13,0.07){\oval(6,3)[t]}

\put(3,2.55){\vector(-1,0){1}}

\put(3,1.55){\vector(1,0){1}}

\put(3,0){$\bullet$}

\put(6,0){$\bullet$}

\put(0,0){$\bullet$}

\put(0,-0.5){$i$}

\put(6,-0.5){$i+1$}

\put(3.5,.8){$C_{i}$}

\put(3,3){$C_{i+1}$}

\put(2,-4.8){$\Large{\sigma_i} $}

\put(13.13,0.07){\oval(6,5)[b]}

\put(13.13,0.07){\oval(6,3)[b]}

\put(13,-2.47){\vector(-1,0){1}}

\put(13,-1.47){\vector(1,0){1}}

\put(13,0){$\bullet$}

\put(16,0){$\bullet$}

\put(10,0){$\bullet$}

\put(10,0.5){$i$}

\put(16,0.5){$i+1$}

\put(12.5,-1){$C_i $}

\put(12.7,-3.2){$C_{i+1}$}

\put(12,-4.8){$\Large{\sigma_{i}^{-1}}$}

\end{picture}

\end{center}
\vskip 3cm
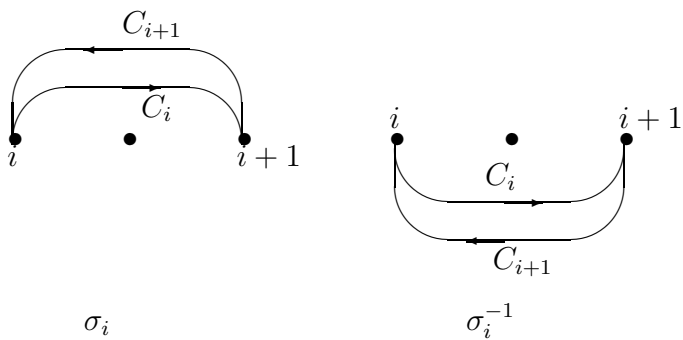
\hspace*{1cm}Figure 1: The interchange of two particles $i$ and $i+1$ along the closed loop $C_{i}C_{i+1}$.\\
\vskip 1cm

Now to look for a unitary one-dimensional representations of $B_N$, we pose $\chi (\sigma_i)=e^{i\phi_i }$ and we have the following constraint
$$\sigma_i \sigma_{i+1}\sigma_i =\sigma_{i+1}\sigma_i \sigma_{i+1}$$
from the Feynman propagator $$K=\sum\limits_{\al\in B_N}\chi(\al)K^\al $$ requiring $\chi (\sigma_i)\chi (\sigma_j)=\chi (\sigma_i\sigma_j)$. The representations $\chi(\al)$ are the weights of different classes $\al$ and the sum runs over all classes. $K^\al$ denotes the integral over all paths in the class $\al$. The constraint $
\sigma_i \sigma_{i+1}\sigma_i = \sigma_{i+1}\sigma_i \sigma_{i+1}$ requires that $\phi_i =\phi_{i+1}$. Thus it is customary to write
\beq
\chi (\sigma_i)=e^{-i\nu\pi },\phantom{~~~}\chi (\sigma_i ^{-1})=e^{i\nu\pi },\phantom{~~~~~~~~~~~~~~~}\nu\in\lbrack 0,2)
\eeq
with $\nu$ is called statistical parameter. If $\nu=0$, the particles are bosons that obey Bose-Einstein statistics and if $\nu=1$ the particles are fermions with Fermi-Dirac statistics.\\

After reviewing in brief the planar system and its statistics, we give in what follows its associated symmetry interpolating between bosonic and fermionic symmetries.

\section{Exotic Particles Algebra}
\hspace*{.3in}First, we briefly recall the non-commutative geometry. Its most simple example consists of the geometric space described by non-commutative hermitian operator coordinates $x_{i}$, and by considering the non-commutative momentum operators $p_i=i\partial_{x_i}$ $(\partial_{x_i}$ the corresponding derivative of $x_i$). These operators satisfy the following algebra
\beq
\bra{cc}
\lbrack x_{i} ,x_{j} \rbrack = i\theta\epsilon_{ij},\phantom{~~~~~}
\lbrack p_i,p_j \rbrack =-i\theta^{-1}\epsilon_{ij},\phantom{~~~~~}\lbrack p_i,x_j \rbrack =i\delta_{ij}\\\\
\lbrack p_i ,t\rbrack =0=\lbrack x_i ,t\rbrack ,\phantom{~~~~~}\lbrack p_i,\partial_{t}\rbrack =0=\lbrack x_i,\partial_{t}\rbrack,
\era
\eeq
with $t$ the physical time and $\partial_{t}$ its corresponding derivative.\\

By considering two-dimensional harmonic oscillator which can be decomposed into one-dimensional oscillators. So, it is known that the algebra (2) allows to define, for each dimension, the representation of annihilation and  creation operators as follows
\beq
\bra{ll}
a_{i}&=\sqrt{\frac{\mu\om}{2}}(x_i +\frac{i}{\mu\om} p_i ) \\
a_{i}^{\da}&=\sqrt{\frac{\mu\om}{2}}(x_i -\frac{i}{\mu\om}  p_i ) .\\
\era
\eeq
with $\mu$ is the mass and $\om$ the frequency. These operators satisfy $$\lbrack a_i ,a^{\da}_i \rbrack =1,$$ defining the Heisenberg algebra. In the simultanuously non-commutative space-space and non-commutative momentum-momentum, the bosonic statistics should be maintained; i.e, the operators $a_{i}^{\da}$ and $a_{j}^{\da}$ are commuting for $i\ne j$. Thus, the deformation parameter $\te$ is required to satisfy the condition $$\te=-(\frac{1}{\mu\om})^2\te^{-1}.$$

To find out an algebra describing the planar system we start by introducing the non-commutative geometry depending on the statistical parameter $\nu\in\bf R\rm$. The fundamental algebra is defined by the coordinates $x_i$ and the momentum $p_i$ satisfying\\

{\bf Proposition 1}
\beq
\bra{cc}
\lbrack x_i ,x_j \rbrack_\chi = i\theta \epsilon_{ij},\phantom{~~~~~} \lbrack p_i ,p_j \rbrack_\chi =- i\theta (\mu\om)^{2}\epsilon_{ij},\phantom{~~~~~}\lbrack p_i ,x_j \rbrack_ =i\delta_{ij} \\\\
\lbrack p_i ,t\rbrack =0=\lbrack x_i ,t\rbrack ,\phantom{~~~~~}\lbrack p_i,\partial_{t}\rbrack =0=\lbrack x_i,\partial_{t}\rbrack
\era
\eeq
and by straightforward calculations we obtain
\beq
\bra{lr}
\lbrace x_i ,p_j\rbrace_\chi =-i\de_{ij}+C_{ji},&
\lbrace p_i ,x_j \rbrace_\chi =i\de_{ij}+D_{ji},
\era
\eeq
where $C_{ji}=(1-\chi)p_j x_i$ and $D_{ji}=(1-\chi)x_j p_i$ and the second deformation parameter $\chi$ is given by\\

{\bf Definition 1}
\beq
\chi=e^{\pm i\nu\pi},
\eeq
where $\pm$ sign indicates the two rotation directions on two-dimensional space. $\te$ in (4) is a non-commutative parameter depending on statistical parameter $\nu$ as we will see later and the notation $[x,y]_q =xy-qyx$.\\

Then, we introduce an operator $\xi_i$ acting on the momentum direction in the phase-space. We assume that $\xi_i$ satisfies the following commutation relation\\

{\bf Proposition 2}
\beq
\lbrack\xi_i ,x_{j}\rbrack =0\phantom{~~~~~}\forall i,j.
\eeq
In this case, we define the annihilation and the creation operators by\\

{\bf Definition 2}
\beq
\bra{ll}
b_{i}^-&= \sqrt{\frac{\mu\om}{2}}(x_i +\frac{i}{\mu\om} \xi_i p_i ) \\ \\
b^+_{i}&= \sqrt{\frac{\mu\om}{2}}(x_i -\frac{i}{\mu\om} \xi^{-1}_i p_i ) ,
\era
\eeq
with $\xi_i $ is defined in terms of statistical parameter $\nu$ and an operaor $K_i $ which could be a function of the number operator $N$\\

{\bf Definition 3}
\beq
\xi_i =e^{i\nu\pi K_i },
\eeq
with $K_i$ an arbitrary operator.

The non-commutative geometry defined by (4-5) leads to a deformed Heisenberg algebra generated by the operators (8) and defined by the following commutation relations
\beq
\bra{ccc}
\lbrack b^-_{i},b^+_{j} \rbrack_\chi = \frac{1}{2}(\xi_i +\xi^{-1}_j ) \de_{ij}+i\frac{\mu\om}{2}\te(I+\xi_i \xi^{-1}_j)\epsilon_{ij} -\frac{i}{2}(\xi_j^{-1}C_{ji}-\xi_i D_{ji}),\\ \\
\lbrack b^+_{i},b^+_{j} \rbrack_\chi = \frac{1}{2}(\xi^{-1}_j -\xi^{-1}_i ) \de_{ij}+ i\frac{\mu\om}{2}\te(I-\xi^{-1}_i \xi^{-1}_j ) \epsilon_{ij} -\frac{i}{2}(\xi_j^{-1}C_{ji}+\xi_i^{-1}D_{ji}),\\ \\
\lbrack b^-_{i},b^-_{j} \rbrack_\chi = \frac{1}{2}(\xi_i -\xi_j ) \de_{ij}+i\frac{\mu\om}{2}\te(I-\xi_i \xi_j )\epsilon_{ij}+\frac{i}{2}(\xi_j C_{ji}+\xi_i D_{ji}).
\era
\eeq
with $I$ is the identity.\\

To be consistent with the hermiticity of the operators $x_i$ and $p_i$ the non-commutative geometry (4) leads to the fact that $\te$ is an operator satisfying $$\theta^\dagger=\chi^{-1}\theta$$ and we suggest the following definition\\

{\bf Definition 4}
\beq
\te =\nu(1+\chi)I
\eeq

We remark that the algebra (10) is a deformed version of Heisenberg algerbra satisfied by the operators given in (3). This new algebra describes the anyonic system for arbitrary statistical parameter $\nu$.\\

Another important point is that the obtained deformed Heisenberg algerbra (10) is interpolating between two extremes depending on  the statistical parameter $\nu$. We know that, in three or more dimensions, $\nu$ takes the values 0 or 1 and in two dimensions $\nu$ is arbitrary real number. The latter case characterizing exotic particles has already discussed above. In the case of three or more dimensions if $\nu=0$ we get $\chi=1$, $\xi_i =\xi^{-1}_i =I$, $\te=0$ and $C_{ji}=0=D_{ji}$, so the commutation relations in the algebra (10) becomes
\beq
\bra{lcr}
\lbrack b^-_{i},b^+_{j} \rbrack = \de_{ij},&
\lbrack b^+_{i},b^+_{j} \rbrack = 0,&
\lbrack b^-_{i},b^-_{j} \rbrack = 0.
\era
\eeq
These relations define the bosonic algebra and this is one extreme. Then the next interesting remark is that the algebra (10) will not have fermionic algebra as extreme for $\nu=1$, $\chi=-1$, $\te=0$ and $C_{ji}\ne 0\ne D_{ji}$ but we get a deformed fermionic algebra as second extreme defined by
\beq
\bra{ccc}
\lbrace b^-_{i},b^+_{j} \rbrace = \frac{1}{2}(e^{i\pi K_i} +e^{-i\pi K_j} ) \de_{ij} -\frac{i}{2}(e^{-i\pi K_j}C_{ji}-e^{i\pi K_i} D_{ji}),\\ \\
\lbrace b^+_{i},b^+_{j} \rbrace = \frac{1}{2}(e^{-i\pi K_j} -e^{-i\pi K_i} ) \de_{ij}-\frac{i}{2}(e^{-i\pi K_j}C_{ji}+e^{-i\pi K_i}D_{ji}),\\ \\
\lbrace b^-_{i},b^-_{j} \rbrace = \frac{1}{2}(e^{i\pi K_i} -e^{i\pi K_j} ) \de_{ij}+\frac{i}{2}(e^{i\pi K_j} C_{ji}+e^{i\pi K_i} D_{ji}).
\era
\eeq

The main result we get from this investigation is that from exotic particles algebra we get the bosonic algebra as extreme this means that our system is originally gotten by exciting a bosonic system in two-dimensional space. On other hand, we remark that for arbitrary operator $K_i$ if $\nu=1$ we get a deformed fermionic algebra as a second extreme. It is known in the literature that anyons are interpolating between bosons and fermions and the statistical parameter $\nu$ equals to 1 describes fermionic system, but in our case, it is different and our system doesn't have any thing to do with fermions originally but it could be related to something else as deformed fermions which are known in the literature as quionic particles or $k_i$-fermions, $k_i$ integer number introduced as a deformation parameter, and these kinds of particles are not physical particles. Consequently, in physicswise, our system has just one extreme which is a bosonic system with the statitical parameter $\nu=0$.

\section{Conclusion}
\hspace{.3in}To conclude, let us say, we started by giving a short review on exotic particles and then we assumed that the fundamental algebra satisfied by the non-commuting coordinates is depending on the statistical parameter $\nu$ characterizing the quasi-particles. Basing on this proposed algebra we defined an annihilation and creation operators in two-dimensional space. Then, by a straightforward calculation we found that the new operators are generators of a deformed Heisenberg algebra describing exotic particles. For arbitrary $\xi_i$ which is introduced to define the "exotic" annihilation and creation operators the two extremes of the exotic particles algebra (10) will be the bosonic algebra and the deformed fermionic one. We also get the same result if $\xi_i$ is unitary and $K_i$ is hermitian as a special case in which the operator $b_i^+$ becomes a complex conjugate of $b_i^-$. 

\section*{knowledgements}
The author would like to thank the Abdus Salam ICTP for the hospitality during the visit in which a part of this work was done.

\end{document}